\documentclass[twocolumn]{aastex7}

\usepackage{graphicx}
\usepackage{amsmath}
\usepackage{hyperref}
\usepackage{longtable}
\usepackage{booktabs}

\newcommand{\te}{t_{\rm E}}
\newcommand{\thetae}{\theta_{\rm E}}

\newcommand{\pie}{\pi_{\rm E}}

\newcommand{\dl}{D_{\rm L}}

\def\e{{\rm E}}

\definecolor{brown}{rgb}{0.59, 0.29, 0.0}
\definecolor{darkgreen}{rgb}{0.0, 0.42, 0.24}
\definecolor{darkblue}{rgb}{0.01, 0.31, 0.59}
\definecolor{darkblue}{rgb}{0.0, 0.25, 0.42}
\definecolor{blue}{rgb}{0.0,0.0,1.0}
\definecolor{green}{rgb}{0.0,1.0,0.0}

\begin{document}

\title{ KMT-2025-BLG-0975Lb and KMT-2025-BLG-1160Lb: Two Uranus-Mass Planets Beyond the Snow Line Discovered by Microlensing}
\shorttitle{Two Uranus-Mass Microlensing Planets}


\author{Cheongho Han}
\affiliation{Department of Physics, Chungbuk National University, Cheongju 28644, Republic of Korea}
\email{cheongho@astroph.chungbuk.ac.kr}
\author{Chung-Uk Lee}
\affiliation{Korea Astronomy and Space Science Institute, Daejon 34055, Republic of Korea}
\email{leecu@kasi.re.kr}
\author{Andrzej Udalski} 
\affiliation{Astronomical Observatory, University of Warsaw, Al.~Ujazdowskie 4, 00-478 Warszawa, Poland}
\email{udalski@astrouw.edu.pl} 
\author{Andrew Gould}
\affiliation{Department of Astronomy, Ohio State University, 140 West 18th Ave., Columbus, OH 43210, USA}
\email{gould.34@osu.edu}
\collaboration{14}{(Leading authors)}
\author{Michael D. Albrow}   
\affiliation{University of Canterbury, Department of Physics and Astronomy, Private Bag 4800, Christchurch 8020, New Zealand}
\email{michael.albrow@canterbury.ac.nz}
\author{Sun-Ju Chung}
\affiliation{Korea Astronomy and Space Science Institute, Daejon 34055, Republic of Korea}
\email{sjchung@kasi.re.kr}

\author{Youn Kil Jung}
\affiliation{Korea Astronomy and Space Science Institute, Daejon 34055, Republic of Korea}
\affiliation{University of Science and Technology, Daejeon 34113, Republic of Korea}
\email{younkil21@gmail.com}
\author{Kyu-Ha~Hwang}
\affiliation{Korea Astronomy and Space Science Institute, Daejon 34055, Republic of Korea}
\email{kyuha@kasi.re.kr}
\author{Yoon-Hyun Ryu}
\affiliation{Korea Astronomy and Space Science Institute, Daejon 34055, Republic of Korea}
\email{yhryu@kasi.re.kr}
\author{Yossi Shvartzvald}
\affiliation{Department of Particle Physics and Astrophysics, Weizmann Institute of Science, Rehovot 76100, Israel}
\email{yossishv@gmail.com}
\author{In-Gu Shin}
\affiliation{Department of Astronomy, Westlake University, Hangzhou 310030, Zhejiang Province, China}
\email{ingushin@gmail.com}
\author{Jennifer C. Yee}
\affiliation{Center for Astrophysics $|$ Harvard \& Smithsonian 60 Garden St., Cambridge, MA 02138, USA}
\email{jyee@cfa.harvard.edu}
\author{Weicheng Zang}
\affiliation{Department of Astronomy, Westlake University, Hangzhou 310030, Zhejiang Province, China}
\email{zangweicheng@westlake.edu.cn}
\author{Hongjing Yang}
\affiliation{Department of Astronomy, Westlake University, Hangzhou 310030, Zhejiang Province, China}
\email{yanghongjing@westlake.edu.cn}
\author{Doeon Kim}
\affiliation{Department of Physics, Chungbuk National University, Cheongju 28644, Republic of Korea}
\email{qso21@hanmail.net}
\author{Dong-Jin Kim}
\affiliation{Korea Astronomy and Space Science Institute, Daejon 34055, Republic of Korea}
\email{keaton03@kasi.re.kr}
\author{Byeong-Gon Park}
\affiliation{Korea Astronomy and Space Science Institute, Daejon 34055, Republic of Korea}
\email{bgpark@kasi.re.kr}
\author{Richard W. Pogge}
\affiliation{Department of Astronomy, Ohio State University, 140 West 18th Ave., Columbus, OH 43210, USA}
\email{pogge.1@osu.edu}
\collaboration{20}{(KMTNet Collaboration)}
\author{Przemek Mr{\'o}z}
\affiliation{Astronomical Observatory, University of Warsaw, Al.~Ujazdowskie 4, 00-478 Warszawa, Poland}
\email{pmroz@astrouw.edu.pl}
\author{Micha{\l} K. Szyma{\'n}ski}
\affiliation{Astronomical Observatory, University of Warsaw, Al.~Ujazdowskie 4, 00-478 Warszawa, Poland}
\email{msz@astrouw.edu.pl}
\author{Jan Skowron}
\affiliation{Astronomical Observatory, University of Warsaw, Al.~Ujazdowskie 4, 00-478 Warszawa, Poland}
\email{jskowron@astrouw.edu.pl}
\author{Rados{\l}aw Poleski} 
\affiliation{Astronomical Observatory, University of Warsaw, Al.~Ujazdowskie 4, 00-478 Warszawa, Poland}
\email{radek.poleski@gmail.co}
\author{Igor Soszy{\'n}ski}
\affiliation{Astronomical Observatory, University of Warsaw, Al.~Ujazdowskie 4, 00-478 Warszawa, Poland}
\email{soszynsk@astrouw.edu.pl}
\author{Pawe{\l} Pietrukowicz}
\affiliation{Astronomical Observatory, University of Warsaw, Al.~Ujazdowskie 4, 00-478 Warszawa, Poland}
\email{pietruk@astrouw.edu.pl}
\author{Szymon Koz{\l}owski} 
\affiliation{Astronomical Observatory, University of Warsaw, Al.~Ujazdowskie 4, 00-478 Warszawa, Poland}
\email{simkoz@astrouw.edu.pl}
\author{Krzysztof A. Rybicki}
\affiliation{Astronomical Observatory, University of Warsaw, Al.~Ujazdowskie 4, 00-478 Warszawa, Poland}
\affiliation{Department of Particle Physics and Astrophysics, Weizmann Institute of Science, Rehovot 76100, Israel}
\email{krybicki@astrouw.edu.pl}
\author{Patryk Iwanek}
\affiliation{Astronomical Observatory, University of Warsaw, Al.~Ujazdowskie 4, 00-478 Warszawa, Poland}
\email{piwanek@astrouw.edu.pl}
\author{Krzysztof Ulaczyk}
\affiliation{Department of Physics, University of Warwick, Gibbet Hill Road, Coventry, CV4 7AL, UK}
\email{kulaczyk@astrouw.edu.pl}
\author{Marcin Wrona}
\affiliation{Astronomical Observatory, University of Warsaw, Al.~Ujazdowskie 4, 00-478 Warszawa, Poland}
\affiliation{Villanova University, Department of Astrophysics and Planetary Sciences, 800 Lancaster Ave., Villanova, PA 19085, USA}
\email{mwrona@astrouw.edu.pl}
\author{Mariusz Gromadzki}          
\affiliation{Astronomical Observatory, University of Warsaw, Al.~Ujazdowskie 4, 00-478 Warszawa, Poland}
\email{marg@astrouw.edu.pl}
\author{Mateusz J. Mr{\'o}z} 
\affiliation{Astronomical Observatory, University of Warsaw, Al.~Ujazdowskie 4, 00-478 Warszawa, Poland}
\email{mmroz@astrouw.edu.pl}
\collaboration{100}{(The OGLE Team)}
\correspondingauthor{\texttt{leecu@kasi.re.kr}}

\begin{abstract}
We present the analysis of two planetary microlensing events, KMT-2025-BLG-0975 and KMT-2025-BLG-1160, 
discovered during the 2025 Galactic bulge microlensing season through high-cadence survey observations.
 In both events, short-duration anomalies near the peaks of the lensing light curves reveal the presence 
of planetary companions.  Light-curve modeling yields planet-to-host mass ratios of $q = 8.6 \times 10^{-4}$ 
for KMT-2025-BLG-0975 and $1.3 \times 10^{-4}$ for KMT-2025-BLG-1160.  For KMT-2025-BLG-0975, finite-source 
effects are detected, enabling a measurement of the angular Einstein radius, whereas only a lower limit on 
this quantity is obtained for KMT-2025-BLG-1160. We estimate the physical parameters of the lens systems 
through Bayesian analyses constrained by the measured microlensing observables.  
The results indicate that the planetary companions have masses of
$M_{\rm p}=29.8^{+50.5}_{-16.0}~M_\oplus$ for KMT-2025-BLG-0975Lb and
$25.4^{+15.5}_{-14.1}~M_\oplus$ for KMT-2025-BLG-1160Lb.
Both planets have masses comparable to that of Uranus.
The host stars are inferred to be a low-mass M dwarf with a mass of
$M_{\rm h}=0.10^{+0.18}_{-0.06}~M_\odot$ for KMT-2025-BLG-0975L and
a late K dwarf with a mass of
$M_{\rm h}=0.58^{+0.35}_{-0.32}~M_\odot$ for KMT-2025-BLG-1160L.
The projected planet--host separations are
$a_\perp=0.81^{+0.10}_{-0.11}$~au for KMT-2025-BLG-0975Lb and
$a_\perp=2.56^{+0.48}_{-0.71}$~au and
$3.29^{+0.61}_{-0.92}$~au for the inner and wide solutions,
respectively, of KMT-2025-BLG-1160Lb.
In both systems, the planets are located beyond the expected snow-line distances of their hosts, placing 
them in the cold ice-giant regime. These discoveries add to the growing sample of Uranus-mass planets 
detected by microlensing and demonstrate that cold ice giants can form and survive around host stars 
spanning a broad range of masses. The results further highlight the increasing sensitivity of modern 
high-cadence microlensing surveys to low-mass planets beyond the snow line.
\end{abstract}


\section{Introduction} \label{sec:one}

Uranus-mass planets, occupying the mass range between super-Earths and gas giants, represent 
a critical transition regime in the framework of planet formation theory. In the core accretion
paradigm, planetary cores form through the accumulation of solid material within protoplanetary
disks \citep{Pollack1996, Ida2004, Mordasini2009}. Once a core attains a critical mass of 
roughly 10 $M_\oplus$, it begins to accrete gas from the surrounding nebula. Whether this gas 
accretion proceeds to runaway growth, forming a Jovian planet, or halts prematurely due to 
disk dissipation or inefficient cooling determines the planet’s final nature \citep{Lee2015, 
Lambrechts2017}.  Uranus-mass planets therefore embody the outcome of failed or incomplete gas 
accretion, providing a natural laboratory for studying the balance between solid and gaseous 
growth phases. Understanding their frequency and physical characteristics provides key 
constraints on the efficiency of core formation and the timescales of disk evolution across 
a range of stellar environments \citep{Mordasini2012, Suzuki2016, Pascucci2018}.

\begin{deluxetable}{lllllll}
\tablewidth{0pt}
\tablecaption{event coordinates and ID correspondence. \label{table:one}}
\tablehead{
\multicolumn{1}{c}{KMTNet}                           &
\multicolumn{1}{c}{$(\alpha, \delta)_{\rm J2000}$}   &
\multicolumn{1}{c}{$(l, b)$}                         &
\multicolumn{1}{c}{OGLE}                       
}
\startdata
 KMT-2025-BLG-0975  &   (17:50:17.84, -30:35:15.00)  &   (-0\fdg8895, -1\fdg7222)   &  \nodata               \\
 KMT-2025-BLG-1160  &   (17:55:26.51, -30:43:29.71)  &   (-0\fdg4463, -2\fdg7454)   &  OGLE-2025-BLG-0941    \\
\enddata
\end{deluxetable}

Gravitational microlensing is uniquely suited to detect such planets, particularly those 
orbiting beyond the snow line, where ices condense and the core accretion process is 
expected to operate most efficiently.  Unlike radial velocity and transit methods, which 
are biased toward short-period, close-in planets, microlensing is sensitive to planets in 
wide orbits around both luminous and faint stars, including M dwarfs and even brown dwarfs 
(e.g., \citealt{Han2013}; \citealt{Jung2018}) and white dwarfs (e.g., \citealt{Blackman2021}).  
This sensitivity makes microlensing the only technique currently capable of probing the 
population of cold Uranus-mass planets located several astronomical units from their hosts.  
Each detection provides an essential data point linking theoretical models of disk evolution 
to the actual occurrence of subcritical cores in the outer regions of planetary systems.

Demographic studies of microlensing planets have shown that intermediate-mass planets 
with masses comparable to those of Neptune and Uranus are common beyond the snow line 
and outnumber Jupiter-mass gas giants \citep{Sumi2010, Gould2006, Suzuki2016, Udalski2018, 
Zang2025}.  This observational trend suggests that runaway gas accretion is not a universal 
outcome of the core accretion process, and that many protoplanetary cores halt their growth 
before reaching the critical mass required for giant planet formation.  The population of 
Uranus-mass planets therefore serves as a key anchor in the planetary mass function between 
the domains of super-Earths and gas giants, providing empirical evidence for a smooth yet 
nontrivial transition in formation efficiency with increasing planetary mass \citep{Suzuki2018, 
Poleski2021}.

Moreover, microlensing detections extend our view of planet formation beyond the Solar
neighborhood. They probe systems located throughout the Galactic disk and bulge, around 
stars of various masses and metallicities. The discovery of Uranus-mass planets in such 
diverse environments demonstrates that ice-giant formation is a robust and widespread 
outcome of planetary evolution, not confined to solar-type stars or specific disk conditions. 
Consequently, Uranus-mass planets observed through microlensing provide a population-level 
perspective on the universality and diversity of planet formation processes across the 
Milky Way, thereby bridging the gap between theoretical predictions and the observed 
architecture of planetary systems.

Looking ahead, the advent of extremely large telescopes (ELTs), in particular
the 39-m European ELT (EELT), will make it possible to transform a large fraction
of microlensing planet detections from mass-ratio measurements into direct
planet-mass measurements (e.g., \citealt{Batista2015, Bennett2015}), typically
5--10 years after the event occurs \citep{Gould2022}. This means that planets
discovered today will contribute to a detailed planet-mass function in the
ice-giant regime relatively soon after EELT first light, which is expected
around 2030.

Despite the scientific importance of Uranus-mass planets, the number of microlensing 
detections remains limited, with only about three hundred microlensing planets discovered 
to date and only a portion of these being Uranus-mass analogs.  As a result, the current 
sample is still insufficient to draw strong statistical constraints on planet formation 
mechanisms or to assess how formation efficiency varies with stellar type and Galactic 
environment. Therefore, it is essential to increase the number of microlensing detections 
through continued high-cadence ground-based surveys and upcoming space-based missions 
such as the Nancy Grace Roman Space Telescope \citep{Penny2019} to improve sensitivity 
to cold, intermediate-mass planets beyond the snow line. Expanding the sample of such 
planets will provide the statistical foundation needed to place robust constraints on 
planet formation pathways and to establish a comprehensive demographic picture of ice 
giants across the Milky Way.

In this paper, we report the discovery of two microlensing planets with masses comparable 
to that of Uranus: KMT-2025-BLG-0975Lb and KMT-2025-BLG-1160Lb. For these events, the 
planetary signals manifest as short-lived anomalies near the peaks of the lensing light 
curves. The excellent temporal coverage provided by the high-cadence surveys enabled the 
unambiguous detection of these signals and facilitated robust modeling of the lensing 
events.

\section{Observations and data} \label{sec:two}

The events were discovered during the 2025 Galactic bulge microlensing season by the 
high-cadence microlensing surveys KMTNet and OGLE, which continuously monitor the 
Galactic bulge in search of gravitational microlensing events. The equatorial and 
Galactic coordinates of the source stars are listed in Table~\ref{table:one}.

KMT-2025-BLG-1160 was observed by both KMTNet and OGLE, whereas KMT-2025-BLG-0975 was 
monitored exclusively by KMTNet. Following the standard microlensing convention of 
adopting the designation assigned by the discovering survey, we refer to the events 
throughout this paper by their KMTNet names.

The KMTNet observations were conducted using three identical 1.6~m telescopes located 
at the Cerro Tololo Inter-American Observatory in Chile (KMTC), the South African 
Astronomical Observatory in South Africa (KMTS), and the Siding Spring Observatory in 
Australia (KMTA). Each telescope is equipped with a $4~{\rm deg}^2$ camera consisting 
of four $9{\rm k}\times9{\rm k}$ CCDs, enabling nearly continuous, high-cadence monitoring 
of the Galactic bulge \citep{Kim2016}. Observations were obtained primarily in the $I$ band, 
with occasional $V$-band images acquired for source-color measurements.

The OGLE observations were carried out using the 1.3~m Warsaw Telescope at Las Campanas 
Observatory in Chile, equipped with a $1.4~{\rm deg}^2$ mosaic CCD camera \citep{Udalski2015}. 
Similar to KMTNet, most observations were taken in the $I$ band, supplemented by a smaller 
number of $V$-band images.

The photometric data were reduced using the pipelines developed by the individual survey 
teams.  The KMTNet data were processed using a customized pipeline based on the difference 
image analysis (DIA) method of \citet{Albrow2009}, while the OGLE photometry was obtained 
using the OGLE-IV DIA pipeline \citep{Wozniak2000, Udalski2003}. To improve the photometric 
precision of the light curve, the KMTNet data were additionally re-reduced using the 
optimized photometry pipeline of \citet{Yang2024}.  The photometric uncertainties were 
renormalized following the procedure of \citet{Yee2012}. For each data set, the error 
bars were adjusted so that the cumulative distribution of $\chi^2$ as a function of 
magnification is approximately linear and the resulting $\chi^2$ per degree of freedom 
is close to unity.

\begin{figure*}[t]
\centering
\includegraphics[width=14.5cm]{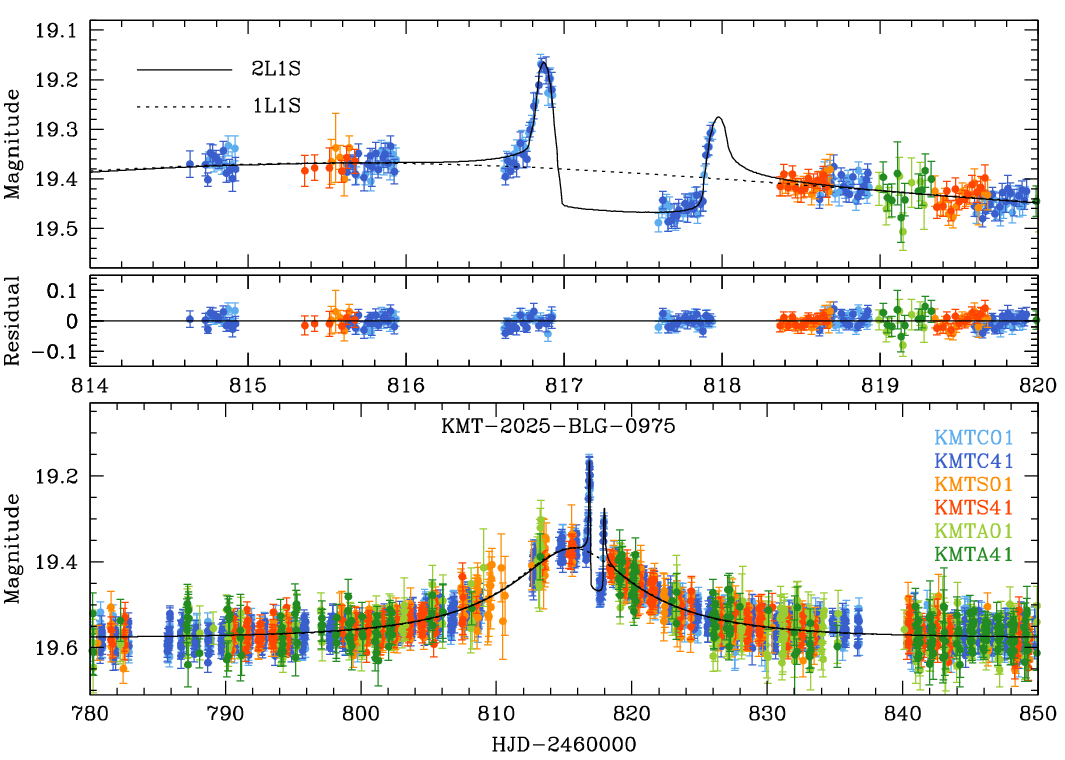}
\caption{
Lensing light curve of KMT-2025-BLG-0975. The bottom panel shows the full light curve, while 
the upper panels present enlarged views of the peak region and the residuals from the best-fit 
planetary model. The dotted and solid curves overlaid on the data represent the best-fit 
single-lens single-source (1L1S) and binary-lens single-source (2L1S) models, respectively.
}
\label{fig:one}
\end{figure*}

\section{Lensing light curve modeling} \label{sec:three}

The two lensing events exhibit broadly similar anomaly characteristics. First, the anomalies 
are short-lived, with durations of $\Delta t \lesssim 1$ day in both events. Second, they 
occur near the peaks of their respective light curves.

Despite these similarities, the events also display notable differences. Their peak magnifications 
differ substantially: KMT-2025-BLG-0975 reached a moderate peak magnification of $A_{\rm max}
\sim5.8$, whereas KMT-2025-BLG-1160 attained a very high peak magnification of $A_{\rm max}>100$. 
In addition, the anomaly morphologies are markedly different. KMT-2025-BLG-1160 exhibits a purely 
negative deviation from the underlying single-lens light curve, whereas KMT-2025-BLG-0975 displays 
a combination of positive and negative deviations.

Short-duration anomalies in microlensing light curves are often indicative of planetary 
perturbations \citep{Gould1992}, although they can also arise from a faint companion to 
the source star rather than from a planetary companion to the lens \citep{Gaudi1998,Gaudi2004}.  
We do not consider the binary-source interpretation further because a binary-source companion 
can only increase the observed magnification, producing positive deviations relative to the 
standard single-source light curve \citep{Gaudi1998,Gaudi2004}, whereas the anomalies in both 
events include negative deviations.

Having established that the observed anomalies are unlikely to originate from binary-source 
companions, we modeled the events using binary-lens single-source (2L1S) models. A standard 
single-lens single-source (1L1S) light curve is characterized by three parameters: the time 
of closest lens--source approach, $t_0$, the lens--source separation at that time, $u_0$, and 
the event timescale, $t_{\rm E}$. For a binary lens, three additional parameters are required 
to describe the lens configuration: the projected separation between the lens components, $s$, 
their mass ratio, $q=M_2/M_1$, and the angle, $\alpha$, between the binary-lens axis and the 
source trajectory. When the source passes close to or crosses a caustic, finite-source effects 
become important. These effects are parameterized by the normalized source radius, 
$\rho=\theta_*/\thetae$, where $\theta_*$ and $\thetae$ denote the angular radii of the source 
star and the Einstein ring, respectively.

The modeling procedure was designed to identify the set of lensing parameters that best 
reproduces the observed light curves.  We first conducted a grid search over the binary-lens 
parameters $(s,q)$ while optimizing the remaining parameters using a Markov Chain Monte Carlo 
(MCMC) algorithm.  The initial search covered a broad parameter space, spanning $-1.0 < \log s 
\leq 1.0$ and $-6.0 < \log q \leq 1.0$, sampled on a $70 \times 70$ grid, to explore both 
stellar- and planetary-mass companions as possible origins of the anomalies. Based on the 
local minima found in this initial search, we progressively narrowed the parameter ranges 
and increased the grid resolution to refine the candidate solutions. The local minima for 
each event are described in detail in the following subsections.  The candidate solutions 
were then refined through full MCMC optimization over all lensing parameters, and their 
relative merits were assessed by comparing the corresponding $\chi^2$ values. The analyses 
of KMT-2025-BLG-0975 and KMT-2025-BLG-1160 are presented separately below.

\begin{deluxetable}{lll}
\tablewidth{\columnwidth}
\tablecaption{Lensing parameters of KMT-2025-BLG-0975. } \label{table:two}
\tablehead{
\multicolumn{1}{c}{Parameter}        &
\multicolumn{1}{c}{Close}            &
\multicolumn{1}{c}{Wide}    
}
\startdata
 $\chi^2$             &  $7351.7           $   &  7618.8\\
 $t_0$ (HJD$^\prime$) &  $815.621 \pm 0.053$   &  $815.537 \pm 0.0549$  \\
 $u_0$                &  $0.323 \pm 0.023  $   &  $0.410 \pm 0.022   $  \\
 $\te$ (days)         &  $13.02 \pm 0.57   $   &  $10.94 \pm 0.45    $  \\
 $s$                  &  $0.8424 \pm 0.0098$   &  $1.2341 \pm 0.0132 $  \\
 $q$ ($10^{-4}$)      &  $8.57 \pm 0.31    $   &  $17.66 \pm 1.47    $  \\
 $\alpha$ (rad)       &  $1.165 \pm 0.013  $   &  $4.322 \pm 0.013   $  \\
 $\rho$ ($10^{-3}$)   &  $6.06 \pm 0.43    $   &  $3.21 \pm 0.38     $  \\
\enddata
\tablecomments{HJD$^\prime = {\rm HJD} - 2460000$.}
\end{deluxetable}

\subsection{KMT-2025-BLG-0975} \label{sec:three-one}

Figure~\ref{fig:one} shows the lensing light curve of KMT-2025-BLG-0975, constructed from 
the combined KMTNet observations obtained at the three network observatories. The event 
reached its peak magnification on 2025 May 20, corresponding to the abridged Heliocentric 
Julian Date ${\rm HJD}^\prime \equiv {\rm HJD} - 2460000 \simeq 816$, with a moderate peak 
magnification of $A_{\rm max} \sim 5.8$. The source star, with a baseline magnitude of 
$I_{\rm base}=19.58$, is located in the overlap region of the KMTNet prime fields BLG01 
and BLG41, which were monitored with a combined cadence of 0.25 hr.  The line of sight 
toward the event is subject to substantial extinction, with $A_I=2.49$, estimated from 
the Galactic bulge extinction map of \citet{Gonzalez2012}.

Inspection of the light curve reveals a short-lived anomaly near the event peak. The upper panel 
of Figure~\ref{fig:one} presents a zoomed-in view of the anomalous region. The dotted curve 
represents the 1L1S model obtained by fitting the light curve after excluding the anomalous 
data. The anomaly is densely covered by KMTC observations obtained over two consecutive nights, 
corresponding to ${\rm HJD}^\prime = 816$ and 817. The data from the first night exhibit a positive 
deviation from the 1L1S model, while those from the second night show a sequence of negative and 
positive deviations. The abrupt changes in magnification, together with the presence of both 
positive and negative deviations, indicate that the anomaly was produced by caustic crossings 
induced by a lens companion.

\begin{figure}[t]
\includegraphics[width=\columnwidth]{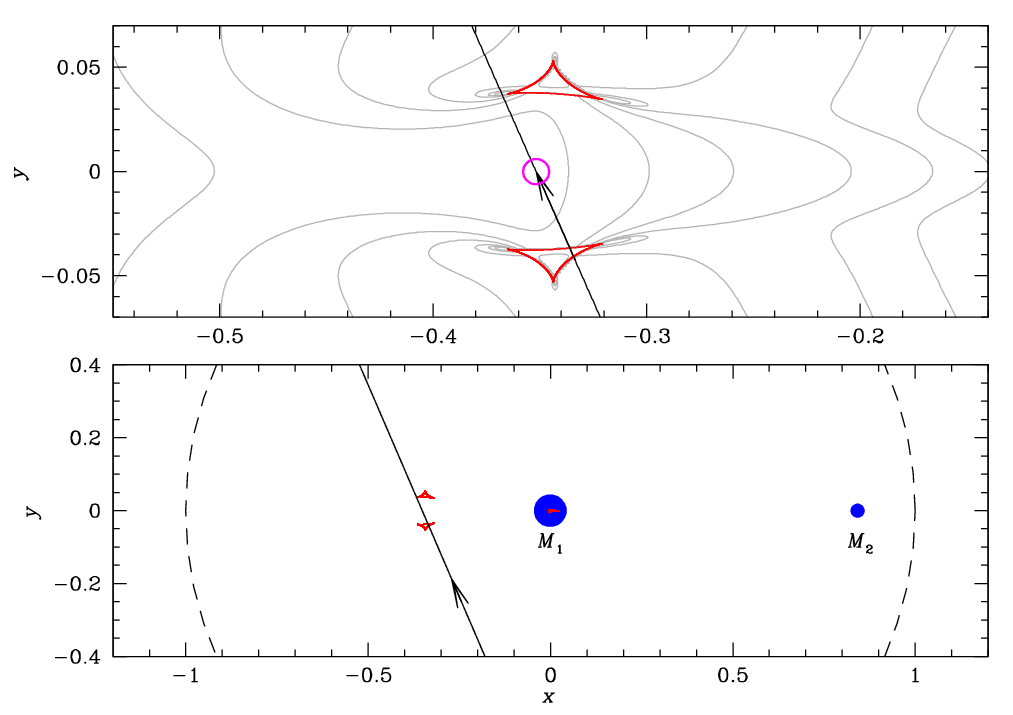}
\centering
\caption{
Configuration of the lens system for KMT-2025-BLG-0975. The lower panel shows the complete 
caustic topology, including both the central and planetary caustics, while the upper panel 
provides a magnified view of the planetary caustics. In each panel, the red cuspy curves 
represent the caustics, and the arrowed curve indicates the source trajectory. The dashed 
circle of unit radius centered at the origin denotes the Einstein ring. The cyan circle on 
the source trajectory marks the source star, with its radius scaled relative to the caustic 
structure.  The two blue filled dots, labeled $M_1$ and $M_2$, indicate the positions of the 
host and planetary lens components, respectively.  The gray curves surrounding the caustics 
are equi-magnification contours.
}
\label{fig:two}
\end{figure}

Motivated by these caustic-related features, we modeled the light curve using a 2L1S
configuration. The modeling yielded a unique planetary solution, whose model light curve 
is shown as the solid curve in Figure~\ref{fig:one}. The best-fit model parameters are 
$(s, q) \sim (0.84, 8.6 \times 10^{-4})$, with an event timescale of $\te \sim 13$~days. 
The full set of lensing parameters is presented in Table~\ref{table:two}.  The characteristic 
anomaly, consisting of two well-resolved caustic crossings, completely breaks the close--wide 
degeneracy. The corresponding wide solution is disfavored by $\Delta\chi^2 = 267.1$. For 
comparison, the lensing parameters of the wide solution are also listed in Table~\ref{table:two}.  
Given that the event timescale is typical of lensing events produced by low-mass stars, the 
small mass ratio indicates that the companion is planetary in nature, with a mass substantially 
lower than that of a Jupiter-mass giant planet. The caustic-crossing features also tightly 
constrain the normalized source radius, yielding $\rho \sim 6.1 \times 10^{-3}$.

\begin{figure*}[t]
\centering
\includegraphics[width=14.5cm]{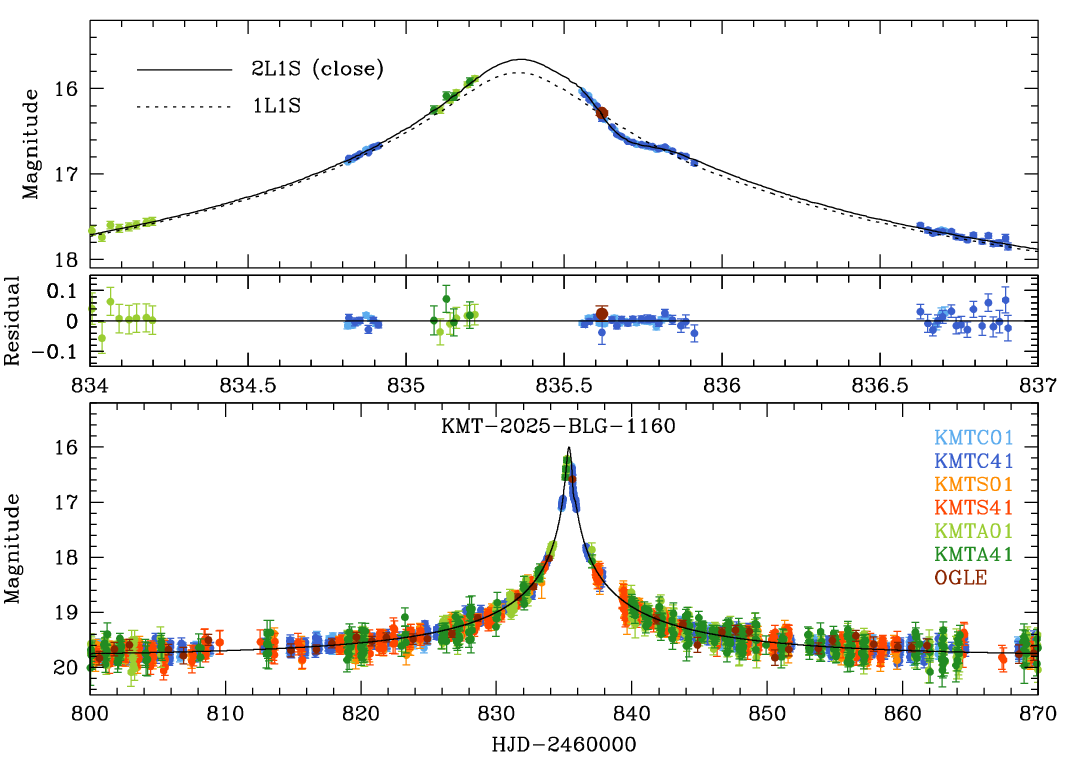}
\caption{
Light curve of the lensing event KMT-2025-BLG-1160.  Among the two degenerate 2L1S 
solutions, the close solution is shown as the representative model.  The notation is 
identical to that used in Fig.~\ref{fig:one}.
}
\label{fig:three}
\end{figure*}

The lens-system configuration corresponding to the best-fit solution is shown in Figure~\ref{fig:two}. 
Because the projected planet--host separation is smaller than the angular Einstein radius ($s<1$), 
the planet induces a central caustic located near the host and a pair of planetary caustics 
situated on the side opposite the planet with respect to the host. The planetary caustics are 
displaced from the host by approximately $s-1/s \sim -0.34$ \citep{Han2006}. The lower panel 
displays the full caustic topology, including both the central and planetary caustics, while 
the upper panel presents an enlarged view of the planetary caustics.

The geometry indicates that the anomaly was generated as the source successively traversed 
the two planetary caustics. The source first crossed the right flank of the lower planetary 
caustic, producing the first anomaly feature, and subsequently crossed the left flank of the 
upper planetary caustic, generating the second feature. The negative deviation observed 
between the two anomaly features occurred when the source passed through the demagnification 
region located between the two planetary caustics.

\subsection{KMT-2025-BLG-1160} \label{sec:three-two}

The light curve of the microlensing event KMT-2025-BLG-1160 is presented in Figure~\ref{fig:three}. 
The event occurred on a source with a baseline magnitude of $I_{\rm base}=19.70$ and is located 
in a field with an $I$-band extinction of $A_I=1.73$. The source lies within the overlapping 
region of the KMTNet prime fields BLG01 and BLG41, which are monitored with a combined cadence 
of 0.25~hr. The event was also independently detected by the OGLE survey and designated 
OGLE-2025-BLG-0941. The event reached its peak magnification on 2025 June 8 (${\rm HJD}^{\prime}\sim 835$), 
attaining a very high magnification of $A_{\rm max}\sim133$.

\begin{deluxetable}{lll}
\tablewidth{0pt}
\tablecaption{Lensing parameters of KMT-2025-BLG-1160.} \label{table:three}
\tablehead{
\multicolumn{1}{c}{Parameter}        &
\multicolumn{1}{c}{Close}            &
\multicolumn{1}{c}{Wide}             
}
\startdata
 $\chi^2$               &  $4828.4             $   &  $4828.7             $  \\  
 $t_0$ (HJD$^\prime$)   &  $835.3578 \pm 0.0021$   &  $835.3572 \pm 0.0025$  \\
 $u_0$ ($10^{-3}$)      &  $7.50 \pm 0.23      $   &  $7.38 \pm 0.28      $  \\
 $\te$ (days)           &  $25.17 \pm 0.56     $   &  $25.55 \pm 0.614    $  \\
 $s$                    &  $0.8717 \pm 0.0087  $   &  $1.1205 \pm 0.0165  $  \\
 $q$ ($10^{-4}$)        &  $1.32 \pm 0.20      $   &  $1.18 \pm 0.22      $  \\
 $\alpha$ (rad)         &  $0.5146 \pm 0.0073  $   &  $0.5132 \pm 0.0086  $  \\
 $\rho$ ($10^{-3}$)     &  $< 2                $   &  $< 2                $  \\
\enddata
\end{deluxetable}

\begin{figure}[t]
\includegraphics[width=\columnwidth]{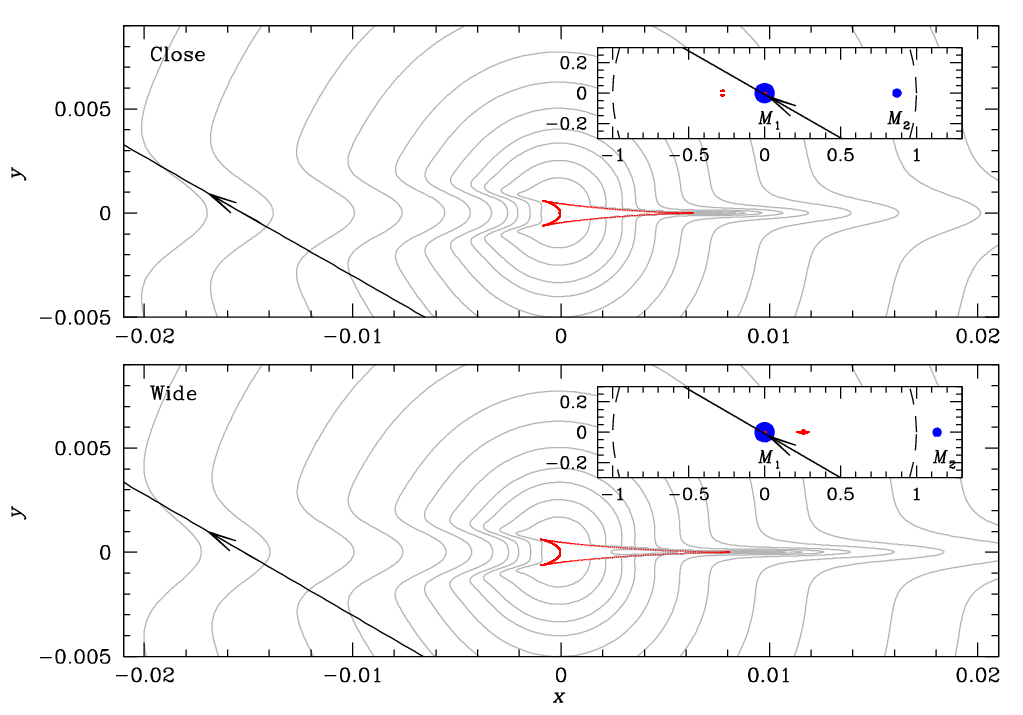}
\centering
\caption{
Configuration of the lens system for KMT-2025-BLG-1160. The upper and lower panels 
correspond to the close and wide solutions, respectively. The insets display the 
full lens geometry, including the central and planetary caustics, the positions of 
the lens components, and the Einstein ring. The gray curves denote contours of 
constant magnification.
}
\label{fig:four}
\end{figure}

Close inspection of the peak region reveals a short-term anomaly. The upper panel of 
Figure~\ref{fig:three} presents a zoomed-in view of the peak, showing that the anomaly 
manifests as a negative deviation from the underlying 1L1S model. Because such short-duration 
negative anomalies are commonly indicative of planetary perturbations \citep{Han2025}, we 
modeled the event using a 2L1S interpretation.

The analysis yields two planetary solutions: one with $(s,q)\sim(0.87,1.3\times10^{-4})$ 
and the other with $(s,q)\sim(1.18,1.2\times10^{-4})$. The former corresponds to a 
configuration in which the projected planet--host separation is smaller than the angular 
Einstein radius ($s<1$) and is therefore referred to as the ``close'' solution, while the 
latter corresponds to a configuration with a separation larger than the Einstein radius ($s>1$) 
and is referred to as the ``wide'' solution.  These two configurations are a classic manifestation 
of the close--wide degeneracy commonly encountered in planetary microlensing events 
\citep{Griest1998}. The degeneracy is extremely severe, with the close solution favored by 
only $\Delta\chi^2=0.3$. The lensing parameters of the two solutions are nearly identical 
except for the separation parameter, which satisfies the relation $s_{\rm close}\sim 1/s_{\rm wide}$, 
as expected for the close--wide degeneracy. The model curve for the close solution is overlaid 
on the data in Figure~\ref{fig:three}, and the full set of lensing parameters for both solutions 
is presented in Table~\ref{table:three}.

The geometric configurations of the two solutions are shown in Figure~\ref{fig:four}, with 
the upper and lower panels corresponding to the close and wide solutions, respectively. Despite 
the difference in the planet--host separation, the two configurations are geometrically very 
similar. In both cases, the anomaly was produced as the source traversed a region of negative 
magnification excess located on the rear side of the central caustic induced by the planet. 
Because the source passed at a relatively large distance from the caustic, finite-source effects 
are only weakly constrained. As a result, the normalized source radius cannot be measured reliably, 
and only an upper limit of $\rho_{\rm max}\sim2\times10^{-3}$ can be placed.

\begin{figure}[t]
\includegraphics[width=\columnwidth]{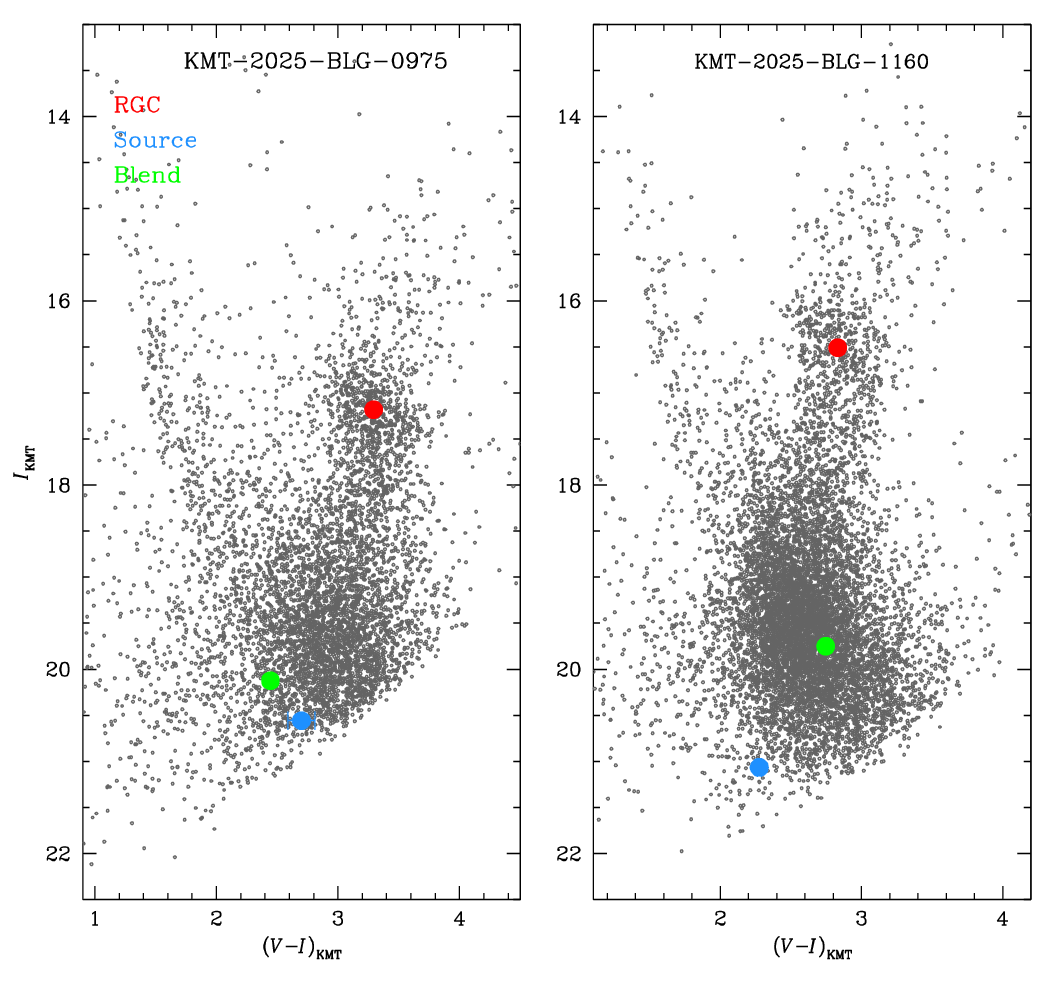}
\centering
\caption{
Locations of the source, centroid of red giant clump (RGC), and blend in the
instrumental color-magnitude diagram constructed using KMTC images. 
}
\label{fig:five}
\end{figure}

\section{Angular Einstein radius} \label{sec:four}

The mass and distance of a lens are constrained by key lensing observables, including the event
timescale, angular Einstein radius, and microlens parallax. The event timescale is routinely
determined from light-curve modeling, and the microlens parallax can also be measured for 
long-timescale events that exhibit noticeable parallax-induced deviations. In contrast, the 
angular Einstein radius is not directly obtained from the modeling but is inferred from the 
combination of the normalized source radius and the angular source radius ($\theta_*$) through 
the relation
\begin{equation}
\theta_{\rm E} = {\theta_* \over \rho}.
\label{eq1}
\end{equation}
\hskip-4pt
Therefore, estimating $\theta_*$ is essential for determining $\theta_{\rm E}$.

To estimate the angular source radius, we identified the source type using its color and 
magnitude, as outlined below.  First, we derived the instrumental color and magnitude of the 
source, $(V-I, I)$, by fitting the model light curves to the observed data in each passband. 
The photometric measurements used for this analysis were obtained from light curves processed 
with the pyDIA photometry code \citep{Albrow2017}.  The code extracts the differential and 
reference fluxes from the template images, allowing the total source flux in the $I$ and $V$ 
bands to be reconstructed in instrumental units for each passband.  Next, we located the source 
on the color-magnitude diagram (CMD) constructed from neighboring field stars using KMTC images. 
The CMD was generated using the same pyDIA photometry to ensure consistent calibration between 
the source and field stars. We then derived the de-reddened color and magnitude of the source, 
$(V-I, I)_0$, by referencing the centroid of the red giant clump (RGC), following the method 
of \citet{Yoo2004}. The intrinsic RGC parameters, $(V-I, I)_{\rm RGC}$, were adopted from the 
calibrations of \citet{Bensby2013} and Table 1 of \citet{Nataf2013}. Finally, the de-reddened 
color and magnitude of the source were estimated as
\begin{equation}
(V-I, I)_0 = 
(V-I, I)_{0,{\rm RGC}} - 
\Delta(V-I, I),   
\label{eq2}
\end{equation}
\hskip-4pt
where $\Delta(V-I, I)=(V-I, I) - (V-I, I)_{\rm RGC}$ denotes the offset between the source 
and the RGC centroid in the instrumental CMD.

Figure~\ref{fig:five} shows the locations of the source stars in the CMDs for the individual 
events, together with the RGC centroids and the blends. The derived values of $(V-I, I)$, 
$(V-I, I)_{\rm RGC}$, $(V-I, I)_{{\rm RGC},0}$, and $(V-I, I)_0$ are listed in 
Table~\ref{table:four}, along with the corresponding spectral types of the source stars. The 
intrinsic source colors indicate that both sources are F-type main-sequence stars according 
to the empirical color--spectral-type calibration of \citet{Pecaut2013}. The blend positions 
are shown in Figure~\ref{fig:five} for completeness and correspond to the blend fluxes derived 
from the light-curve modeling. They are not used in the source-color calibration, and therefore 
only the source and RGC parameters are listed in Table~\ref{table:four}.

\begin{deluxetable}{lll}[t]
\tabletypesize{\footnotesize}
\tablewidth{0pt}
\tablecaption{Source parameters, angular Einstein radius, and relative lens-source proper motion. } \label{table:four}
\tablehead{
\multicolumn{1}{c}{Parameter}               &
\multicolumn{1}{c}{KMT-2025-BLG-0975}       &
\multicolumn{1}{c}{KMT-2025-BLG-1160}       
}
\startdata
 $(V-I)$                      &   $2.699 \pm 0.111 $  &   $2.276 \pm 0.015 $    \\
 $I$                          &   $20.557 \pm 0.018$  &   $21.066 \pm 0.002$    \\
 $(V-I, I)_{\rm RGC}$         &   $(3.292, 17.181) $  &   $(2.832, 16.510) $    \\
 $(V-I, I)_{{\rm RGC},0}$     &   $(1.060, 14.496) $  &   $(1.060, 14.470) $    \\
 $(V-I)_0$                    &   $0.467 \pm 0.118 $  &   $0.504 \pm 0.043 $    \\
 $I_0$                        &   $17.872 \pm 0.027$  &   $19.025 \pm 0.020$    \\
  Spectral type               &    F2V -- F8V         &    F3V -- F6V           \\
 $\theta_*$ ($\mu$as)         &   $0.651 \pm 0.090 $  &   $0.396 \pm 0.032 $    \\
 $\thetae$ (mas)              &   $0.108 \pm 0.017 $  &   $> 0.20          $    \\
 $\mu$ (mas/yr)               &   $3.017 \pm 0.468 $  &   $> 2.1           $    \\
\enddata
\end{deluxetable}

The angular radius of the source star was determined from its estimated color and magnitude. 
For this purpose, we applied the empirical relation between $\theta_*$ and $(V-K, I)$ 
established by \citet{Kervella2004}. To utilize this relation, the measured $(V-I)$ color 
was converted into $(V-K)$ color using the color-color relation of \citet{Bessell1988}. 
Adopting the more recent calibration of \citet{Pecaut2013} would increase the estimated 
angular source radii by only 2.2\% and 2.0\% for KMT-2025-BLG-0975 and KMT-2025-BLG-1160, 
respectively, well below the quoted uncertainties.  With the derived $\theta_*$, the angular 
Einstein radius, $\theta_{\rm E}$, was then computed using Eq.~(\ref{eq1}). Combining 
$\theta_{\rm E}$ with the event timescale, we derived the relative lens-source proper motion as
\begin{equation}
\mu = {\thetae \over \te}.
\label{eq3}
\end{equation}
\hskip-4pt
The resulting values of $\theta_*$, $\theta_{\rm E}$, and $\mu$ for the individual events 
are summarized in Table~\ref{table:four}.  For KMT-2025-BLG-1160, in which only an upper 
limit on $\rho$ is constrained, we report the lower limits of $\theta_{\rm E}$ and $\mu$.

\begin{figure}[t]
\includegraphics[width=\columnwidth]{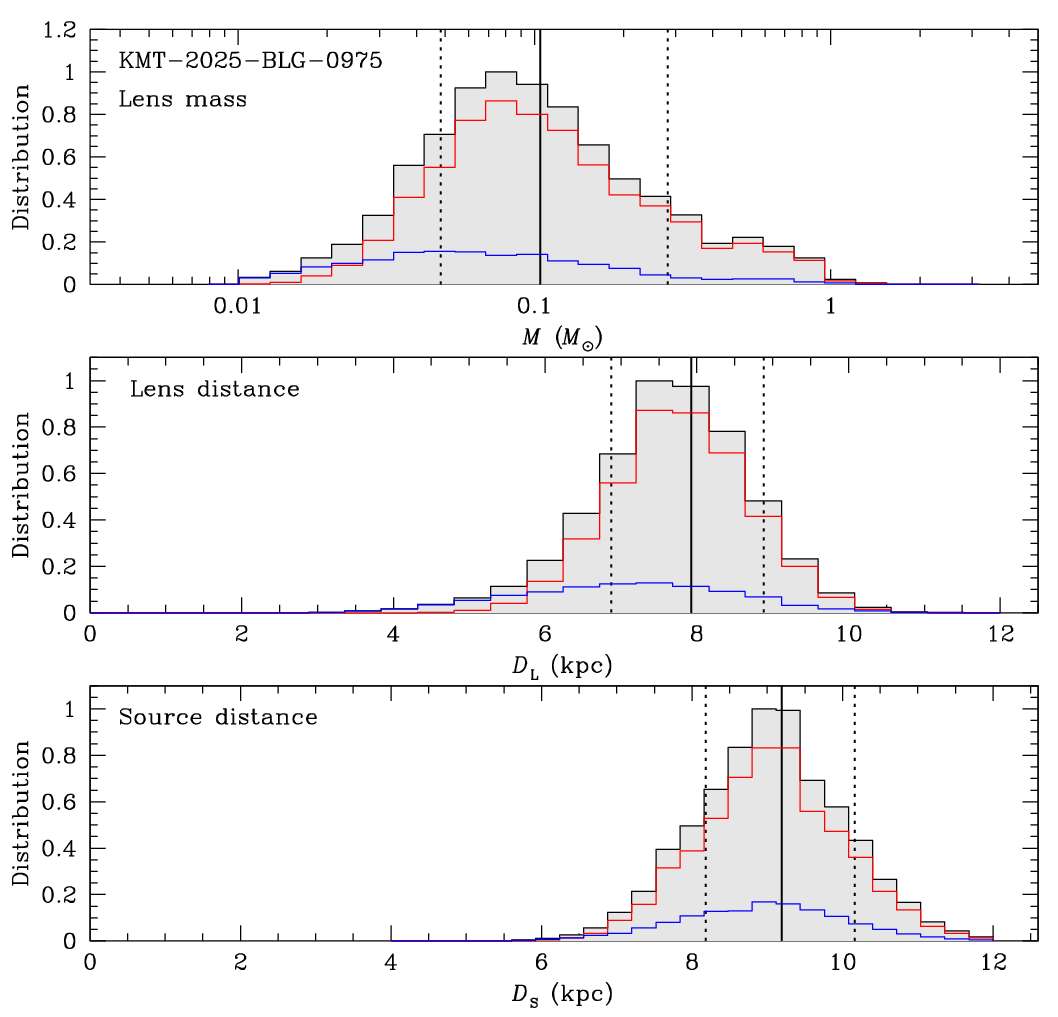}
\centering
\caption{
Bayesian posterior probability distributions of the host-star mass, lens distance, and source 
distance for the microlensing event KMT-2025-BLG-0975.  In each panel, the red and 
blue curves represent the contributions from the disk and bulge lens populations, respectively, 
while the black curve shows their combined distribution. The solid vertical line marks the 
median of the posterior distribution, and the two dotted vertical lines indicate the 1$\sigma$ 
confidence interval.
}
\label{fig:six}
\end{figure}

\section{Planet masses} \label{sec:five}

The physical parameters of a lens are constrained by lensing observables, including 
the event time scale, angular Einstein radius, and microlens parallax ($\pi_{\rm E}$). 
With all the measured observables, the mass and distance of the lens is determined 
using the \citet{Gould2000} relation as
\begin{equation}
M = {\thetae \over \kappa \pie};\qquad
\dl = { {\rm AU} \over \pie\thetae + \pi_{\rm S}}.
\label{eq4}
\end{equation}
\hskip-4pt
Here $\kappa =4G/(c^2{\rm AU}) \simeq 8.14~({\rm mas}/M_\odot)$ and $\pi_{\rm S}$ represents 
the parallax of the source star. This formulation directly links the measurable quantities 
$\theta_{\rm E}$ and $\pi_{\rm E}$ to the physical parameters of the lens system, enabling 
a unique determination of both the lens mass and its distance.

For the analyzed events, the lensing observables were only partially measured. Specifically, 
for KMT-2025-BLG-0975, both the event timescale, $\te$, and the angular Einstein radius, 
$\theta_{\rm E}$, were measured, whereas for KMT-2025-BLG-1160, $\te$ was measured and only 
a lower limit on $\theta_{\rm E}$ could be established. For neither event was the microlens 
parallax, $\pi_{\rm E}$, detected.

\begin{figure}[t]
\includegraphics[width=\columnwidth]{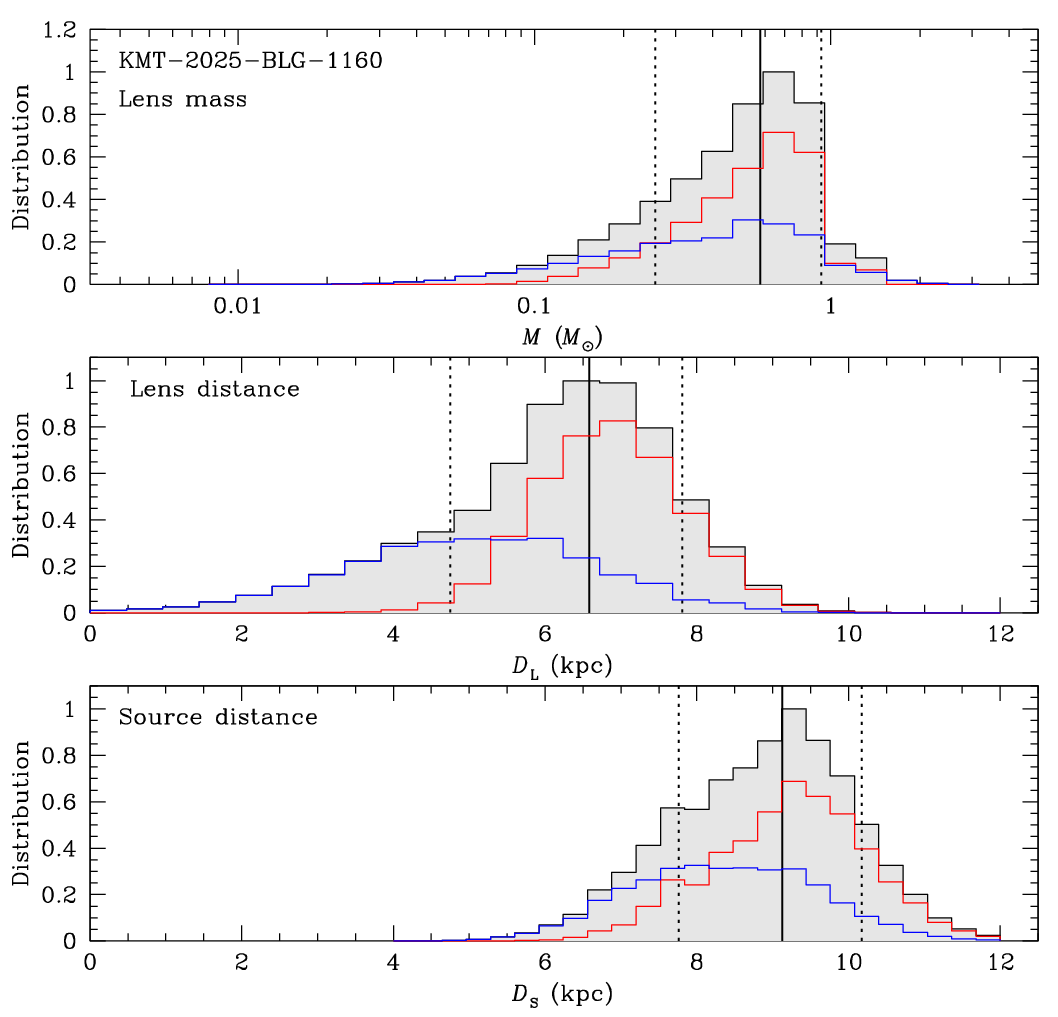}
\centering
\caption{
Posterior distributions of the host-star mass, lens distance, and source distance derived 
from the Bayesian analysis of KMT-2025-BLG-1160.  The curves and vertical lines follow 
the same notation as in Fig.~\ref{fig:six}.
}
\label{fig:seven}
\end{figure}

Given these incomplete observational constraints, the physical parameters of the lens 
systems were estimated through Bayesian analyses. A detailed description of this procedure 
is provided in previous studies (e.g., \citealt{Han2025}) and is therefore only briefly 
summarized here. The analysis combines the measured lensing observables with Galactic 
model priors describing the spatial, kinematic, and mass distributions of potential lens 
populations. Based on these priors, a large ensemble of synthetic microlensing events is 
generated, and each realization is assigned a weight according to its consistency with 
the observed lensing constraints. The posterior probability distributions of the lens 
masses and distances are then derived from the ensemble of weighted events.

\begin{deluxetable}{lll}
\tabletypesize{\small}
\tablewidth{0pt}
\tablecaption{Physical lens parameters.} \label{table:five}
\tablehead{
\multicolumn{1}{c}{Parameter}               &
\multicolumn{1}{c}{KMT-2025-BLG-0975}       &
\multicolumn{1}{c}{KMT-2025-BLG-1160}       
}
\startdata
  $M_{\rm p}$ ($M_\oplus$)  &   $29.80^{+50.46}_{-16.04}$   &  $25.36^{+15.46}_{-14.12}$          \\ [0.8ex]
  \hskip15pt $(M_{\rm U}$)  &   $1.75^{+3.47}_{-1.10}   $   &  $1.74^{+1.06}_{-0.97}   $          \\ [0.8ex]
  $M_{\rm h}$ ($M_\odot$)   &   $0.10^{+0.18}_{-0.06}   $   &  $0.58^{+0.35}_{-0.32}   $          \\ [0.8ex]
  $\dl$ (kpc)               &   $7.923^{+0.96}_{-1.05}  $   &  $6.58^{+1.22}_{-1.83}   $          \\ [0.8ex]
  $a_\perp$ (AU)            &   $0.81^{+0.10}_{-0.11}   $   &  $2.56^{+0.48}_{-0.71}   $ (close)  \\ [0.8ex]
                            &   \nodata                     &  $3.29^{+0.61}_{-0.92}   $ (wide)   \\ [0.8ex]
  $p_{\rm disk}$            &    19\%                       &   41\%                              \\ [0.8ex]
  $p_{\rm bulge}$           &    81\%                       &   59\%                              \\ [0.8ex]
\enddata
\tablecomments{$M_{\rm U}$ denotes the mass of Uranus.}
\end{deluxetable}

The Bayesian posterior distributions of the host-star mass, lens distance, and source 
distance are shown in Figure~\ref{fig:six} for KMT-2025-BLG-0975 and Figure~\ref{fig:seven} 
for KMT-2025-BLG-1160.  The source-distance distributions are included to illustrate the 
relative lens--source locations along the line of sight.

The resulting physical parameters of the lens systems are summarized in Table~\ref{table:five}, 
including the planet mass ($M_{\rm p}$), host-star mass ($M_{\rm h}$), lens distance 
($D_{\rm L}$), and projected planet--host separation ($a_\perp$). Both systems host planets 
with masses comparable to that of Uranus, placing them in the ice-giant regime. However, 
their host stars differ significantly: the host of KMT-2025-BLG-0975 is a low-mass M dwarf, 
while the host of KMT-2025-BLG-1160 is a late K-type dwarf.  Although the posterior distribution 
for KMT-2025-BLG-0975 favors a low-mass M-dwarf host, it extends into the brown-dwarf regime, 
indicating that a brown-dwarf host cannot be ruled out.

Given that the snow-line distance scales approximately as 
$a_{\rm sl}\simeq 2.7~{\rm AU}~(M_{\rm h}/M_\odot)$, 
the measured projected separations are significantly larger than $a_{\rm sl}$ even 
before accounting for projection effects. This indicates that both planets reside well 
beyond the snow line, in the cold outer regions of their planetary systems.

Also presented in the table are the probabilities that the lens belongs to the disk 
($p_{\rm disk}$) or bulge ($p_{\rm bulge}$) population.  For KMT-2025-BLG-0975 the 
lenses are very likely to reside in the bulge, with $p_{\rm bulge} \sim 81\%$.  In the 
case of KMT-2025-BLG-1160, the lens is likewise more likely to be located in the bulge, 
but the probability of it being in the disk is not negligible.

\section{Summary and Conclusion} \label{sec:six}

We have presented the analyses of two planetary microlensing events, KMT-2025-BLG-0975 and 
KMT-2025-BLG-1160, discovered during the 2025 Galactic bulge microlensing season through 
high-cadence survey observations. In both events, short-duration anomalies near the peaks 
of the lensing light curves revealed the presence of planetary companions. By modeling the 
light curves and combining the measured lensing observables with Bayesian analyses, we 
estimated the physical parameters of the lens systems.

For KMT-2025-BLG-0975, the planetary signal was produced by the source successively traversing 
two planetary caustics induced by a companion with a mass ratio of $q=8.6\times10^{-4}$. The 
clear detection of finite-source effects enabled a measurement of the angular Einstein radius, 
$\theta_{\rm E}=0.108\pm0.017$ mas.  The Bayesian analysis indicates that the lens system 
consists of a planet with a mass of $M_{\rm p}=29.8^{+50.5}_{-16.0}~M_\oplus$ orbiting a 
host with a mass of $M_{\rm h}=0.10^{+0.18}_{-0.06}~M_\odot$, which is most likely a low-mass 
M dwarf.

For KMT-2025-BLG-1160, the planetary signal appeared as a short-duration negative perturbation 
near the peak of a high-magnification event with $A_{\rm max}\sim133$. The modeling yielded a 
pair of close--wide degenerate solutions with mass ratios of $q\sim1.2\times10^{-4}$. 
Although finite-source effects were not detected, the Bayesian analysis indicates a 
$M_{\rm p}=25.4^{+15.5}_{-14.1}~M_\oplus$ planet orbiting a host of mass 
$M_{\rm h}=0.58^{+0.35}_{-0.32}~M_\odot$, most likely a late K dwarf.

A notable common characteristic of the two systems is that their planetary companions have 
masses comparable to that of Uranus, placing them in the ice-giant regime. Despite this 
similarity in planetary mass, the host stars differ substantially, spanning a range from a 
low-mass M dwarf to a late K dwarf. In both systems, the projected planet--host separations 
are significantly larger than the expected snow-line distances of their hosts, indicating 
that the planets reside in the cold outer regions of their planetary systems where giant-planet 
cores are expected to form most efficiently.

The two planets reported in this work add to the growing sample of Uranus-mass planets discovered 
by microlensing. Their detections further demonstrate the capability of modern high-cadence 
microlensing surveys to identify low-mass planets beyond the snow line around a wide variety 
of host stars. Continued discoveries of such systems will improve statistical constraints on 
the occurrence rate of cold ice giants and provide important empirical tests of planet-formation 
theories in the regime between terrestrial planets and gas giants.

\begin{acknowledgments}
C.H. was supported by the Chungbuk National University 2025 NUDP program and the National 
Research Foundation of Korea (RS-2025-21073000).
This research has made use of the KMTNet system operated by the Korea Astronomy and Space 
Science Institute (KASI) at three host sites of CTIO in Chile, SAAO in South Africa, and 
SSO in Australia. Data transfer from the host site to KASI was supported by the Korea 
Research Environment Open NETwork (KREONET). This research was supported by KASI under 
the R\&D program (project No. 2026-1-904-01) supervised by the Ministry of Science and ICT.
H.Y. and W.Z. acknowledge support by the National Natural Science Foundation of China 
(Grant No. 12133005). H.Y. acknowledge support by the China Postdoctoral Science Foundation 
(No. 2024M762938).
The OGLE project has received funding from the Polish National Science Centre grant OPUS-28 
2024/55/B/ST9/00447 to AU.
\end{acknowledgments}



\bibliographystyle{aasjournal}
\bibliography{pasp_refs}

@ARTICLE{Pecaut2013,
  author = {Pecaut, Mark J. and Mamajek, Eric E.},
  title = {Intrinsic Colors, Temperatures, and Bolometric Corrections of Pre-main-sequence Stars},
  journal = {The Astrophysical Journal Supplement Series},
  year = {2013},
  volume = {208},
  pages = {9},
  doi = {10.1088/0067-0049/208/1/9}
}

@ARTICLE{Gonzalez2012,
  author = {Gonzalez, O.~A. and Rejkuba, M. and Zoccali, M. and
            Valenti, E. and Minniti, D. and Schultheis, M. and
            Tobar, R. and Chen, B.},
  title = {Reddening and Metallicity Maps of the Milky Way Bulge from VVV and 2MASS.  II.},
  journal = {\aap},
  year = {2012},
  volume = {543},
  pages = {A13},
  doi = {10.1051/0004-6361/201219222}
}

@ARTICLE{Sumi2010,
  author = {Sumi, T. and Bennett, D. P. and Bond, I. A. and Udalski, A. and Batista, V. and 
            Dominik, M. and Fouqu{\'e}, P. and Kubas, D. and Gould, A. and et al.},
  title = {A Cold Neptune-Mass Planet OGLE-2007-BLG-368Lb: Cold Neptunes Are Common},
  journal = {\apj},
  year = {2010},
  volume = {710},
  number = {2},
  pages = {1641},
  doi = {10.1088/0004-637X/710/2/1641}
}

@article{Blackman2021,
  author = {Blackman, J. W. and Beaulieu, J. P. and Bennett, David P. and Danielski, C. and Alard, C. and Cole, A. A. and Vandorou, A. and Ranc, C. and Terry, S. K. and et al.},
  title = {A Jovian Analogue Orbiting a White Dwarf Star},
  journal = {Nature},
  year = {2021},
  volume = {598},
  pages = {272},
  doi = {10.1038/s41586-021-03869-6}
}

@article{Jung2018,
  author = {Jung, Youn Kil and Udalski, Andrzej and Gould, Andrew and Ryu, Yoon-Hyun and Yee, Jennifer C. and Han, Cheongho and Albrow, Michael D. and Lee, Chung-Uk and Kim, Seung-Lee and et al.},
  title = {OGLE-2017-BLG-1522: A Giant Planet around a Brown Dwarf Located in the Galactic Bulge},
  journal = {\aj},
  year = {2018},
  volume = {155},
  pages = {219},
  doi = {10.3847/1538-3881/aabb51}
}

@article{Han2013,
  author = {Han, Cheongho and Jung, Youn Kil and Udalski, Andrzej and Sumi, Takahiro and Gaudi, B. Scott and Gould, Andrew and Bennett, David P. and Tsapras, Y. and Szyma{\'n}ski, M. K. and  et al.},
  title = {Microlensing Discovery of a Tight, Low-mass-ratio Planetary-mass Object around an Old Field Brown Dwarf},
  journal = {The Astrophysical Journal},
  year = {2013},
  volume = {778},
  pages = {38},
  doi = {10.1088/0004-637X/778/1/38}
}

@article{Gould2022,
  author        = {Gould, Andrew},
  title         = {MASADA: From Microlensing Planet Mass-Ratio Function to Planet Mass Function},
  year          = {2022},
  journal       = {arXiv e-prints},
  eid           = {arXiv:2209.12501},
  pages         = {arXiv:2209.12501},
  archivePrefix = {arXiv},
  eprint        = {2209.12501},
  primaryClass  = {astro-ph.EP},
  doi           = {10.48550/arXiv.2209.12501}
}

@article{Bennett2015,
  author  = {Bennett, D. P. and Bhattacharya, A. and Anderson, J. and Bond, I. A. and Anderson, N. and Barry, R. and Batista, V. and Beaulieu, J.-P. and DePoy, D. L. and et al.},
  year    = {2015},
  title   = {Confirmation of the Planetary Microlensing Signal and Star and Planet Mass Determinations for Event OGLE-2005-BLG-169},
  journal = {\apj},
  volume  = {808},
  pages   = {169},
  doi     = {10.1088/0004-637X/808/2/169 }
}

@article{Batista2015,
  author  = {Batista, V. and Beaulieu, J.-P. and Bennett, D. P. and Gould, A. and Marquette, J.-B. and Fukui, A. and Bhattacharya, A.},
  year    = {2015},
  title   = {Confirmation of the OGLE-2005-BLG-169 Planet Signature and Its Characteristics with Lens-Source Proper Motion Detection},
  journal = {\apj},
  volume  = {808},
  pages   = {170},
  doi     = {10.1088/0004-637X/808/2/170}
}

@article{Han2006,
  author  = {Han, C.},
  year    = {2006},
  title   = {Properties of Planetary Caustics in Gravitational Microlensing},
  journal = {\apj},
  volume  = {638},
  pages   = {1080},
  doi     = {10.1086/498937}
}

@article{Gould2006,
  author  = {Gould, A. and Udalski, A. and An, D. and Bennett, D. P. and Zhou, A.-Y. and Dong, S. and Rattenbury, N. J. and Gaudi, B. S. and Yock, P. C. M. and Bond, I. A. and Christie, G. W.},
  year    = {2006},
  title   = {Microlens OGLE-2005-BLG-169 Implies That Cool Neptune-like Planets Are Common},
  journal = {\apj},
  volume  = {644},
  pages   = {L37},
  doi     = {10.1086/505421}
}

@article{Gould2000,
  author  = {Gould, A.},
  year    = {2000},
  title   = {A Natural Formalism for Microlensing},
  journal = {\apj},
  volume  = {542},
  pages   = {785},
  doi     = {10.1086/317037}
}

@article{Kervella2004,
  author  = {Kervella, P. and Th\'evenin, F. and Di Folco, E. and S\'egransan, D.},
  year    = {2004},
  title   = {The angular sizes of dwarf stars and subgiants. Surface brightness relations calibrated by interferometry},
  journal = {\aap},
  volume  = {426},
  pages   = {29},
  doi     = {10.1051/0004-6361:20035930}
}

@article{Griest1998,
  author  = {Griest, K. and Safizadeh, N.},
  year    = {1998},
  title   = {The Use of High-Magnification Microlensing Events in Discovering Extrasolar Planets},
  journal = {\apj},              
  volume  = {500},
  pages   = {37},
  doi     = {10.1086/305729}
}

@article{Penny2019,
  author  = {Penny, M. T. and Gaudi, B. S. and Kerins, E. and Rattenbury, N. J. and Mao, S. and Robin, A. C. and Calchi Novati, S.},
  year    = {2019},
  title   = {Predictions of the WFIRST Microlensing Survey. I. Bound Planet Detection Rates},
  journal = {\apjs},              
  volume  = {241},
  pages   = {3},
  doi     = {10.3847/1538-4365/aafb6}
}

@article{Pascucci2018,
  author  = {Pascucci, I. and Mulders, G. D. and Gould, A. and Fernandes, R.},
  year    = {2018},
  title   = {A Universal Break in the Planet-to-star Mass-ratio Function of Kepler MKG Stars},
  journal = {\apjl},              
  volume  = {856},
  pages   = {L28},
  doi     = {10.3847/2041-8213/aab6ac}
}

@article{Poleski2021,
  author  = {Poleski, R. and Skowron, J. and Mr\'oz, P. and Udalski, A. and Szyma\'nski, M. K. and Pietrukowicz, P. and Ulaczyk, K. and Rybicki, K. and Iwanek, P. and Wrona, M. and Gromadzki, M.},
  year    = {2021},
  title   = {Wide-Orbit Exoplanets are Common. Analysis of Nearly 20 Years of OGLE Microlensing Survey Data},
  journal = {Acta Astron.},              
  volume  = {71},
  pages   = {1},
  doi     = {10.32023/0001-5237/71.1.1}
}

@article{Suzuki2016,
  author  = {Suzuki, D. and Bennett, D. P. and Sumi, T. and Bond, I. A. and Rogers, L. A. and Abe, F. and Asakura, Y. and Bhattacharya, A. and Donachie, M. and Freeman, M.},
  year    = {2016},
  title   = {The Exoplanet Mass-ratio Function from the MOA-II Survey: Discovery of a Break and Likely Peak at a Neptune Mass},
  journal = {\apj},              
  volume  = {833},
  pages   = {145},
  doi     = {10.3847/1538-4357/833/2/145}
}

@article{Suzuki2018,
  author  = {Suzuki, D. and Bennett, D. P. and Ida, S. and Mordasini, C. and Bhattacharya, A. and Bond, I. A. and Donachie, M.},
  year    = {2018},
  title   = {Microlensing Results Challenge the Core Accretion Runaway Growth Scenario for Gas Giants},
  journal = {\apjl},              
  volume  = {869},
  pages   = {L34},
  doi     = {10.3847/2041-8213/aaf577}
}

@article{Lambrechts2017,
  author  = {Lambrechts, M. and Lega, E.},
  year    = {2017},
  title   = {Reduced gas accretion on super-Earths and ice giants},
  journal = {\aap},              
  volume  = {606},
  pages   = {A146},
  doi     = {10.1051/0004-6361/201731014}
}

@article{Lee2015,
  author  = {Lee, E. J. and Chiang, E.},
  year    = {2015},
  title   = {To Cool is to Accrete: Analytic Scalings for Nebular Accretion of Planetary Atmospheres},
  journal = {\apj},              
  volume  = {811},
  pages   = {L41},
  doi     = {10.1088/0004-637X/811/1/41}
}

@article{Pollack1996,
  author  = {Pollack, J. B. and Hubickyj, O. and Bodenheimer, P. and Lissauer, J. J. and Podolak, M. and Greenzweig, Y.},
  year    = {1996},
  title   = {Formation of the Giant Planets by Concurrent Accretion of Solids and Gas},
  journal = {Icarus},              
  volume  = {124},
  pages   = {62},
  doi     = {10.1006/icar.1996.0190}
}

@article{Ida2004,
  author  = {Ida, S. and Lin, D. N. C.},
  year    = {2004},
  title   = {Toward a Deterministic Model of Planetary Formation. I. A Desert in the Mass and Semimajor Axis Distributions of Extrasolar Planets},
  journal = {\apj},              
  volume  = {604},
  pages   = {388},
  doi     = {10.1086/381724}
}

@article{Mordasini2012,
  author  = {Mordasini, C. and Alibert, Y. and Benz, W. and Klahr, H. and Henning, T.},
  year    = {2012},
  title   = {Extrasolar planet population synthesis . IV. Correlations with disk metallicity, mass, and lifetime},
  journal = {\aap},              
  volume  = {541},
  pages   = {A97},
  doi     = {10.1051/0004-6361/201117350 }
}

@article{Mordasini2009,
  author  = {Mordasini, C. and Alibert, Y. and Benz, W.},
  year    = {2009},
  title   = {Extrasolar planet population synthesis. I. Method, formation tracks, and mass-distance distribution},
  journal = {\aap},              
  volume  = {501},
  pages   = {1139},
  doi     = {10.1051/0004-6361/200810301 }
}

@article{Yoo2004,
  author  = {Yoo, J. and DePoy, D. L. and Gal-Yam, A. and Gaudi, B. S. and Gould, A. and Han, C. and Lipkin, Y. and Maoz, D. and Ofek, E. O. and Park, B.-G. and Pogge, R. W. and Udalski, A.},
  year    = {2004},
  title   = {OGLE-2003-BLG-262: Finite-Source Effects from a Point-Mass Lens},
  journal = {\apj},              
  volume  = {603},
  pages   = {139},
  doi     = {10.1086/381241}
}

@article{Albrow2009,
  author  = {Albrow, M. D. and Horne, K. and Bramich, D. M. and Fouqu\'e, P. and Miller, V. R. and 
Beaulieu, J.-P. and Coutures, C. and Menzies, J. and Williams, A.},
  year    = {2009},
  title   = {Difference imaging photometry of blended gravitational microlensing events with a numerical kernel},
  journal = {\mnras},              
  volume  = {397},
  pages   = {2009},
  doi     = {10.1111/j.1365-2966.2009.15098.x}
}

@article{Albrow2017,
  author  = {Albrow, M.},
  year    = {2017},
  title   = {MichaelDAlbrow/pyDIA: Initial Release on Github,Versionv1.0.0},
  journal = {MichaelDAlbrow/pyDIA: Initial Release on Github,Versionv1.0.0, Zenodo},              
  volume  = {},
  pages   = {},
  doi     = {10.5281/zenodo.268049}
}

@article{Bensby2013,
  author  = {Bensby, T. and Yee, J. C.  and Feltzing, S.￼ and Johnson, J. A. and Gould, A. and 
Cohen, J. G. and Asplund, M.  and Mel\'endez, J. and Lucatello, S. ￼ and Han, C. and Thompson, I. and 
Gal-Yam, A. and Udalski, A. ￼and Bennett, D. P. ￼ and Bond, I. A. and Kohei, W. and Sumi, T. and 
Suzuki, D. and Suzuki, K. and Takino, S. and Tristram, P. and Yamai, N. and Yonehara, A.},
  year    = {2013},
  title   = {Chemical evolution of the Galactic bulge as traced by microlensed dwarf and subgiant 
stars. V. Evidence for a wide age distribution and a complex MDF},
  journal = {\aap},              
  volume  = {549},
  pages   = {A247},
  doi     = {10.1051/0004-6361/201220678}
}

@article{Bessell1988,
  author  = {Bessell, M. S. and Brett, J. M.},
  year    = {1988},
  title   = {JHKLM Photometry: Standard Systems, Passbands, and Intrinsic Colors},
  journal = {\pasp},              
  volume  = {100},
  pages   = {1134},
  doi     = {10.1086/132281}
}

@ARTICLE{Gould1992,
  author = {Gould, A. and Loeb, A.},
  title = {Discovering Planetary Systems Through Gravitational Microlenses},
  journal = {ApJ},
  year = {1992},
  volume = {396},
  pages = {104},
  doi    = {10.1086/171700} 
}

@article{Han2025,
  author  = {Han, C. and Bond, I. A. and Jung, Y. K. and Albrow, M. D. and Chung, S.-J. and Gould, A. and Hwang, K.-H. and Lee, C.-U.},
  year    = {2025},
  title   = {MOA-2022-BLG-033Lb, KMT-2023-BLG-0119Lb, and KMT-2023-BLG-1896Lb: Three low mass-ratio microlensing planets detected through dip signals},
  journal = {\aap},              
  volume  = {694},
  pages   = {A90},
  doi     = {10.1051/0004-6361/202452027}
}

@article{Gaudi1998,
  author  = {Gaudi, B. S},
  year    = {1998},
  title   = {Distinguishing Between Binary-Source and Planetary Microlensing Perturbations},
  journal = {\apj},              
  volume  = {506},
  pages   = {533},
  doi     = {10.1086/306256}
}

@article{Gaudi2004,
  author  = {Gaudi, B. S. and Han, C.},
  year    = {2004},
  title   = {The Many Possible Interpretations of Microlensing Event OGLE 2002-BLG-055},
  journal = {\apj},              
  volume  = {611},
  pages   = {528},
  doi     = {10.1086/421971}
}

@article{Yang2024,
  author  = {Yang, H. and Yee, J. C. and Hwang, K.-H. and Qian, Q. and Bond, I. A. and Gould, A. and Hu, Z. and Zhang, J. and Mao, S.},
  year    = {2024},
  title   = {Systematic reanalysis of KMTNet microlensing events, paper I: Updates of the photometry pipeline and a new planet candidate},
  journal = {\mnras},
  volume  = {528},
  pages   = {11},
  doi     = {10.1093/mnras/stad3672},
}

@article{Kim2016,
  author  = {Kim, S.-L. and Lee, C.-U.and Park, B.-G. and Kim, D.-J. and Cha, S.-M. and Lee, Y. and Han, C. and Chun, M.-Y. and Yuk, I.},
  year    = {2016},
  title   = {KMTNET: A Network of 1.6 m Wide-Field Optical Telescopes Installed at Three Southern Observatories},
  journal = {JKAS},              
  volume  = {49},
  pages   = {37},
  doi     = {10.5303/JKAS.2016.49.1.37}
}

@article{Nataf2013,
  author  = {Nataf, David M. and Gould, Andrew and Fouqué, Pascal and Gonzalez, Oscar A. and Johnson, Jennifer A. and Skowron, Jan and Udalski, Andrzej and Szymański, Michal K. and Kubiak, Marcin and Pietrzy\'nski, Grzegorz and Soszy\'nski, Igor and Ulaczyk, Krzysztof and Wyrzykowski, Lukasz and Poleski, Rados andaw},
  year    = {2013},
  title   = {Reddening and Extinction toward the Galactic Bulge from OGLE-III: The Inner Milky Way's RV ~ 2.5 Extinction Curve},
  journal = {\apj},              
  volume  = {769},
  pages   = {88},
  doi     = {10.1088/0004-637X/769/2/88}
}

@article{Udalski2015,
  author  = {Udalski, A. and Szyma\'nski, M. K. and Szyma\'nski, G.},
  year    = {2015},
  title   = {OGLE-IV: Fourth Phase of the Optical Gravitational Lensing Experiment},
  journal = {Acta Astron.},              
  volume  = {65},
  pages   = {1},
  doi     = {10.48550/arXiv.1504.05966}
}

@article{Udalski2018,
  author  = {Udalski, A. and Ryu, Y.-H. and Sajadian, S. and Gould, A. and Mr\'oz, P. and Poleski, R. and Szyma\'nski, M. K. and Skowron, J. and Soszy\'nski, I. and Koz{\l}owski, S.},
  year    = {2018},
  title   = {OGLE-2017-BLG-1434Lb: Eighth q<1×10-4 Mass-Ratio Microlens Planet Confirms Turnover in Planet Mass-Ratio Function},
  journal = {Acta Astron.},              
  volume  = {68},
  pages   = {1},
  doi     = {10.32023/0001-5237/68.1.1}
}

@article{Udalski2003,
  author  = {Udalski, A.},
  year    = {2003},
  title   = {The Optical Gravitational Lensing Experiment. Real Time Data Analysis Systems in the OGLE-III Survey},
  journal = {Acta Astron.},              
  volume  = {53},
  pages   = {291},
  doi     = {10.48550/arXiv.astro-ph/040112}
}

@article{Wozniak2000,
  author  = {Wozniak, P. R.},
  year    = {2000},
  title   = {Difference Image Analysis of the OGLE-II Bulge Data. I. The Method},
  journal = {Acta Astron.},              
  volume  = {50},
  pages   = {421},
  doi     = {10.48550/arXiv.astro-ph/0012143}
}

@article{Yee2012,
  author  = {Yee, J. C. and Shvartzvald, Y. and Gal-Yam, A. and Bond, I. A. and Udalski, A. and Kozłowski, S. and Han, C. and Gould, A. and Skowron, J. and Suzuki, D. and Abe, F. and Bennett, D. P. and Botzler, C. S. and Chote, P. and Freeman, M. and Fukui, A. and Furusawa, K. and Itow, Y. and Kobara, S. and Ling, C. H. and Masuda, K. and Matsubara, Y.},
  year    = {2012},
  title   = {MOA-2011-BLG-293Lb: A Test of Pure Survey Microlensing Planet Detections},
  journal = {\apj},              
  volume  = {755},
  pages   = {102},
  doi     = {10.1088/0004-637X/755/2/102}
}

@article{Zang2025,
  author  = {Zang, W. and Jung, Y. K. and Yee, J. C. and Hwang, K.-H. and Yang, H. and Udalski, A. and Sumi, T. and Gould, A.},
  year    = {2025},
  title   = {Microlensing events indicate that super-Earth exoplanets are common in Jupiter-like orbits},
  journal = {Science},              
  volume  = {388},
  pages   = {400},
  doi     = {10.1126/science.adn6088}
}

\end{document}